\documentclass[aps,prd,groupedaddress,twocolumn,showpacs,nofootinbib]{revtex4}
\usepackage{graphicx,epsfig}
\begin{document}

%\preprint{}

\title{Constraints from primordial black hole formation at the end of inflation}



\author{Amandeep S. Josan}
\email[]{ppxaj1@nottingham.ac.uk}

%\thanks{}

\affiliation{School of Physics and Astronomy, University of Nottingham, University Park, Nottingham, NG7 2RD, UK}

\author{Anne M. Green}
\email[]{anne.green@nottingham.ac.uk}

%\thanks{}

\affiliation{School of Physics and Astronomy, University of Nottingham, University Park, Nottingham, NG7 2RD, UK}


\date{\today}

\begin{abstract}
  Primordial black hole (PBH) abundance limits constrain the
  primordial power spectrum, and hence models of inflation, on scales
  far smaller than those probed by cosmological observations.  Single
  field inflation models which are compatible with all cosmological
  data can have large enough perturbations on small scales to
  overproduce PBHs, and hence be excluded. The standard formulae for
  the amplitude of perturbations do not hold for modes that exit the
  horizon close to the end of inflation however.  We use a modified
  flow analysis to identify models of inflation where the amplitude of
  perturbations on small scales is large. For these models we then
  carry out a numerical evolution of the perturbations and use the PBH
  constraints on the power spectrum to eliminate models which
  overproduce PBHs. Significant PBH formation can occur in models in
  which inflation can continue indefinitely and is ended via a secondary
  mechanism. We demonstrate that PBHs constrain these types
  of inflation models and show that a numerical evaluation of the
  power spectrum decreases  the number of
  otherwise viable models of inflation.
\end{abstract}

%\keywords{}

\maketitle

%\section{}

\section{Introduction}

Primordial black holes (PBHs) can form in the early Universe via the
collapse of large density
perturbations~\cite{Carr:1974nx,Carr:1975qj}. There are tight
constraints on the abundance of PBHs formed from their present day
gravitational effects and the consequences of their evaporation. These
limits can be used to constrain the power spectrum of the primordial
density, or curvature, perturbations.  The PBH constraints on the
curvature power spectrum are fairly weak, being many orders of
magnitude larger than the measurements on cosmological scales. They
do, however, apply over a very wide range of scales  and
therefore provide a useful constraint on models of inflation.  Peiris
and Easther~\cite{Peiris:2008be} have shown that there are single
field inflation models, which are compatible with all cosmological
observations, for which the perturbation amplitude on small scales is
large enough to produce a significant density of PBHs.

For scales which exit the horizon close to the end of inflation the standard
(Stewart-Lyth~\cite{Stewart:1993bc}) formulae for the amplitude
of perturbations do not hold. Leach and Liddle~\cite{Leach:2000yw}
carried out a numerical calculation of the evolution of perturbations 
for a quadratic inflationary potential. They found that the
perturbations on scales which exit the horizon close to the end of
inflation were roughly an order of magnitude larger than predicted by
the Stewart-Lyth formula (see also Ref.~\cite{Bugaev:2008bi}). Therefore 
to fully exploit the power of PBH constraints on inflation models, a 
numerical calculation of the amplitude of perturbations on small scales is
required. It has recently been shown~\cite{Lyth:2005ze,Zaballa:2009xb} that
PBHs can also form on scales which never leave the horizon.  We do not consider 
this possibility here.

In this paper we use a modified flow analysis to identify inflation
models where the perturbations at the end of inflation may be large
enough for primordial black holes to be overproduced. For these models
we carry out a numerical evolution of the primordial perturbations and
use the PBH constraints on the power spectrum to eliminate models
which overproduce PBHs.  We describe the modified flow analysis in
Sec.~\ref{flow} and the evolution of perturbations and the calculation
of the power spectrum in Sec.~\ref{pert}. We apply the primordial
black hole abundance constraints and present our results in
Sec.~\ref{results} and conclude with discussion in Sec.~\ref{discuss}.

\section{Method}
\subsection{Flow equations approach}
\label{flow}

We consider the Hubble slow roll-parameters~\cite{Salopek:1990jq}: 
\begin{eqnarray}
\epsilon_H &=&\frac{m_{{\rm{Pl}}}^2}{4\pi}\left(\frac{H^{\prime}(\phi)}{H(\phi)}\right)^2 \,, \\
^l\lambda_H  & = & \left( \frac{m_{{\rm{Pl}}}^2}{4\pi}\right)^l  \frac{(H^{\prime})^{l-1}}{H^l}
          \frac{{\rm d}^{(l+1)}H}{{\rm d}\phi^{(l+1)}} \hspace{5mm}  l\geq1 \,,
       \label{Hslowrollpara}
\end{eqnarray} 
where $m_{\rm{Pl}}$ is the Planck mass and ${}^\prime$ denotes
differentiation with respect to the scalar field, $\phi$. The flow equations~\cite{Hoffman:2000ue,Kinney:2002qn} encode the variation of the
slow-roll parameters in terms of the number of e-foldings from the end of inflation, $N=\ln{[a(t_{\rm{end}})/a(t)]}$ and provide a method for stochastically 
generating inflation models:
\begin{eqnarray}
\frac{ {\rm d}\epsilon_H}{{\rm d}N} & = & \epsilon_H (\sigma_H+2\epsilon_H) \,,  \\         
\frac{{\rm d} \sigma_H}{{\rm d}N} & = & -5\epsilon_H \sigma_H -12 \epsilon_H^2+2(^2\lambda_H) \,, \\
\frac{{\rm d}(^l \lambda_H )}{{\rm d}N} & = & \left[ \frac{l-1}{2} \sigma_H + (l-2) \epsilon_H \right](^l\lambda_H) +^{l+1}\lambda_H \,,   \\
&& \hspace{50mm} l\ge 1               \nonumber
\label{flowequations}
\end{eqnarray}
where $\sigma_H=2(^1\lambda_H)-4\epsilon_H$ and $^1\lambda_H \equiv \eta_H$.

Following Kinney~\cite{Kinney:2002qn} we randomly chose `initial' values for the slow-roll parameters and $N_{\rm{cos}}$, the number of e-foldings between
cosmological scales exiting the horizon and the end of inflation, in the ranges:
\begin{eqnarray}              \nonumber
N_{\rm{cos}}&=&[40,60] \,, \\  \nonumber
\epsilon_H&=&[0,0.8]  \,, \\   \nonumber
\sigma_H&=&[-0.5,0.5] \,, \\   \nonumber
^2\lambda_H \equiv \xi_H&=&[-0.05,0.05]  \,,\\     \nonumber
^3\lambda_H&=&[-0.005,0.005]  \,, \\   \nonumber
&...& \\
^{M+1}\lambda_H&=&0 \,,
\label{hierarchy}
\end{eqnarray}
truncating the hierarchy at $M=6$.  We then evolve the flow equations
forward in time (${\rm d}N<0$) from $N=N_{\rm{cos}}$ until either $N=0$ or
inflation ends with $\epsilon_H=1$.  In the former case we calculate
the cosmological observables, the spectral index, $n_{\rm s}$, its
running, ${\rm d}  n_{\rm s} /{\rm d} {\rm{ln}} k$, and the scalar to tensor
ratio, $r$, using the initial values of the slow-roll
parameters~\cite{Kinney:2002qn}:
\begin{eqnarray}
n_{\rm s}-1&=& \sigma_H-(5-3C_1)\epsilon_H^2-\frac{1}{4}(3-5C_1)\sigma_H\epsilon_H \nonumber \\
&& +\frac{1}{2}(3-C_1) \xi_H \,, \\
\frac{{\rm d} n_{\rm s}}{{\rm d}{\rm{ln}}{k}}
 &=& -\left(\frac{1}{1-\epsilon_H}\right) \left[2\xi_H-12\epsilon_H^2-5\epsilon_H \sigma_H 
\right. \nonumber \\
&& -\left.\frac{(3-5C_1)}{2}\epsilon_H \xi_H + \frac{(3-C_1)}{4}\sigma_H \xi_H \right]  \,, \\
 r &=&\epsilon_H [1-C_1(\sigma_H+2\epsilon_H)] \,,  
\end{eqnarray}
where $C_1=4(\rm{ln}2+\gamma)-5 \approx 0.0814514$ and $\gamma \approx
0.577$. In the latter case we evolve
the flow equations backward $N_{\rm{cos}}$ e-folds and calculate the
cosmological observables at this point.  In some cases inflation also 
ends when evolving backwards before $N_{\rm{cos}}$ e-folds are achieved. 
These models are incapable of supporting the required amount of inflation 
and are discarded.

Our algorithm differs from that originally proposed by
Kinney~\cite{Kinney:2002qn} in how we handle models chosen from the
initial hierarchy that are destined to
inflate forever, $\epsilon_H \rightarrow 0$, but do not reach this
limit within $N_{\rm{cos}}$ e-foldings.  In the original flow algorithm
in this case the cosmological observables are calculated at the
late-time fixed point i.e. the model is forced to evolve to its
asymptotic limit. In this limit the running of the spectral index is
negligible. Therefore for models which are compatible with the WMAP 7
year measurement of the spectral index, $n_{\rm s}=0.964 \pm
0.012$~\cite{Komatsu:2010fb}, the amplitude of the curvature
perturbations can not be large on any scale and PBHs are never formed
in significant numbers~\cite{Chongchitnan:2006wx}. Following Peiris
and Easther~\cite{Peiris:2008be}, we do not force these models to
evolve to their asymptotic limit but instead terminate them once
$N_{\rm{cos}}$ e-folds of inflation have occurred.  At this point it is
assumed that another mechanism, for example a second-field such as in
hybrid inflation~\cite{Linde:1993cn}, terminates
inflation.  With this treatment some
of these models are consistent with the WMAP measurements of the
spectral index and its running, but have perturbations on small scales
which may be large enough to over-produce PBHs~\cite{Peiris:2008be,Alabidi:2009bk,Kohri:2007qn,Bugaev:2008gw}.

%We note that the detailed properties of inflation models generated by
%the flow equations approach should not be over interpreted.  Firstly
%the flow equations generate models in which the Hubble parameter is a
%polynomial in $\phi$~\cite{Liddle:2003py}. Furthermore alternative
%stochastic methods which use other quantities, for instance
%$\epsilon(\phi)$ or $V(\phi)$, find significantly different
%distributions for the values of $n_{\rm s}$ and $r$ on cosmological
%scales~\cite{Ramirez:2005cy}.  None the less the flow equations
%algorithm allows us to `randomly' generate a large number of models to
%confront with the PBH abundance constraints, using an accurate
%calculation of the primordial perturbations.

\subsection{Perturbation calculation}
\label{pert}

The evolution of inflationary curvature perturbations, ${\cal R}$, is carried out using the Mukhanov variable~\cite{Mukhanov:1990me}, $u = -z {\cal R}$,
where 
\begin{equation}
z= \frac{a}{H} \frac{{\rm d} \phi}{{\rm d} t} \,.
\end{equation}
The Fourier modes, $u_{k}$, evolve according to a
Klein-Gordon equation with a time-varying effective mass:
\begin{equation}
\frac{{\rm d}^2 u_k}{{\rm d}\tau^2}+\left(k^2-\frac{1}{z}\frac{{\rm d}^2z}{{\rm d}\tau^2}\right)u_k=0 \,,
 \label{mukhanovEOMtau}
\end{equation}
where $\tau$ is  conformal time, ${\rm d} \tau \equiv {\rm d} t/a$, and
\begin{eqnarray}
\frac{1}{z}\frac{{\rm d}^2z}{{\rm d}\tau^2} &=& 2a^2H^2[1+\epsilon_H-\frac{3}{2}\eta_H+\epsilon_H^2-2\epsilon_H\eta_H  \nonumber \\
  &&
+\frac{1}{2}\eta_H^2+\frac{1}{2}\xi_H]  \,.
    \label{exactdespite}
\end{eqnarray}
At early times, $\tau_{\rm i}$, when a mode $k$ is well within the horizon, $aH/k  \rightarrow 0$,
the initial condition for $u_{k}(\tau_{i})$, is taken to be the Bunch-Davies vacuum state,
\begin{equation}
u_{k}(\tau_{\rm i}) = \frac{1}{\sqrt{2 k}} \exp{(-i k \tau_{\rm i})} \,.
\label{bd}
\end{equation}
In the superhorizon limit, $k^2 \ll z^{\prime \prime}$,
eq.~(\ref{mukhanovEOMtau}) has a growing mode solution $u_{k} \propto
z$, so that the curvature perturbation ${\cal R}_{k}= |u_{k} /z|$
`freezes out' and becomes constant. The power-spectrum of the
curvature perturbations can thus be calculated as
\begin{equation}
\mathcal{P_\mathcal{R}}(k) \equiv  \frac{k^3}{2 \pi^2}  \left| {\cal R}_{k} \right| ^2 
=\frac{k^3}{2\pi^2} \left| \frac{u_k}{z}\right| ^2 \,.
\label{numericalpowerspectrum}
\end{equation}
Eq.~(\ref{mukhanovEOMtau}) can be solved exactly for the special case
of power-law inflation.  The commonly used Stewart-Lyth formula is
found via a slow-roll expansion around this exact solution~\cite{Stewart:1993bc}:
\begin{equation}
\mathcal{P_R}(k) \approx \frac{[1-(2C+1)\epsilon_H+C\eta_H]^2}{\pi\epsilon_H} \left(\frac{H}{m_{\rm{Pl}}}\right)^2_{k=aH} \,,		\label{stewartlyth}
\end{equation}
where $C=-2+\ln{2}+\gamma \approx -0.729$.  This expression gives the
power spectrum in the asymptotic superhorizon limit, $k/aH \rightarrow
0$, in terms of the Hubble parameter and slow-roll parameters
evaluated at horizon crossing~\cite{Grivell:1996sr}.  It is valid
provided that the slow-roll approximation holds (specifically that the
slow-roll parameters are slowly varying around horizon crossing) and
the asymptotic limit is reached before inflation ends~\cite{Huang:2000bh}.
For modes which exit the horizon close to the end of inflation the 
asymptotic limit will not be reached, and the slow-roll approximation may 
also be violated.
\begin{figure}
\begin{center}
\includegraphics[width=8cm]{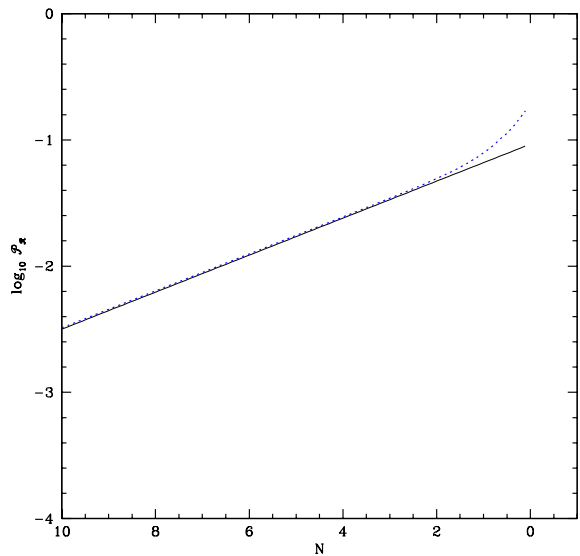}
\caption{The power spectrum of the primordial curvature perturbations
generated during the final few e-folds for an example inflation
model. The black solid line shows the Stewart-Lyth equation while the blue
dotted line is the result of a numerical mode by mode evolution.}
\label{numericalenhancement}
\end{center}
\end{figure}
Leach \& Liddle~\cite{Leach:2000yw} investigated this for a simple quadratic
chaotic inflation model.
%Because the asymptotic regime is not
%adequately reached it is not clear when the perturbation amplitude
%should be evaluated. They considered three epochs: horizon exit during
%inflation, at the end of inflation, and at horizon re-entry after
%inflation.  They found that the amplitude at horizon exit is typically
%significantly greater than at the other two epochs.
They found that for scales that
exit the horizon very close to the end of inflation the power
spectrum is roughly an order of
magnitude larger than that found using the Stewart-Lyth expression.
In other words, analytic calculations can significantly underestimate
the amplitude of perturbations and hence the abundance of PBHs formed.
Therefore, a numerical calculation of the perturbation evolution is
required to accurately compute the primordial power spectrum on the
very smallest scales.

We use a modified version of the Inflation v2
module (written by Lesgourgues \& Valkenburg)~\cite{Lesgourgues:2007aa} 
to carry out an accurate
numerical calculation of the evolution of perturbations.  
Fig.~\ref{numericalenhancement} shows the
power spectrum of curvature perturbations of an example inflation model
generated using the modified horizon flow formalism.  The power
spectrum on large scales is compatible with the WMAP 7 year data,
while the perturbations on small scales are sufficiently large that
PBHs may be over-produced. The Stewart-Lyth calculation is in good
agreement with the numerical calculation until the final few e-folds
of inflation.  On these small scales, the assumptions that are
employed in the Stewart-Lyth calculation break down, and the numerical
calculation finds a significant enhancement of the amplitude of the
perturbations.

\section{Results}
\label{results}

\begin{figure}
\begin{center}
\includegraphics[width=4.1cm]{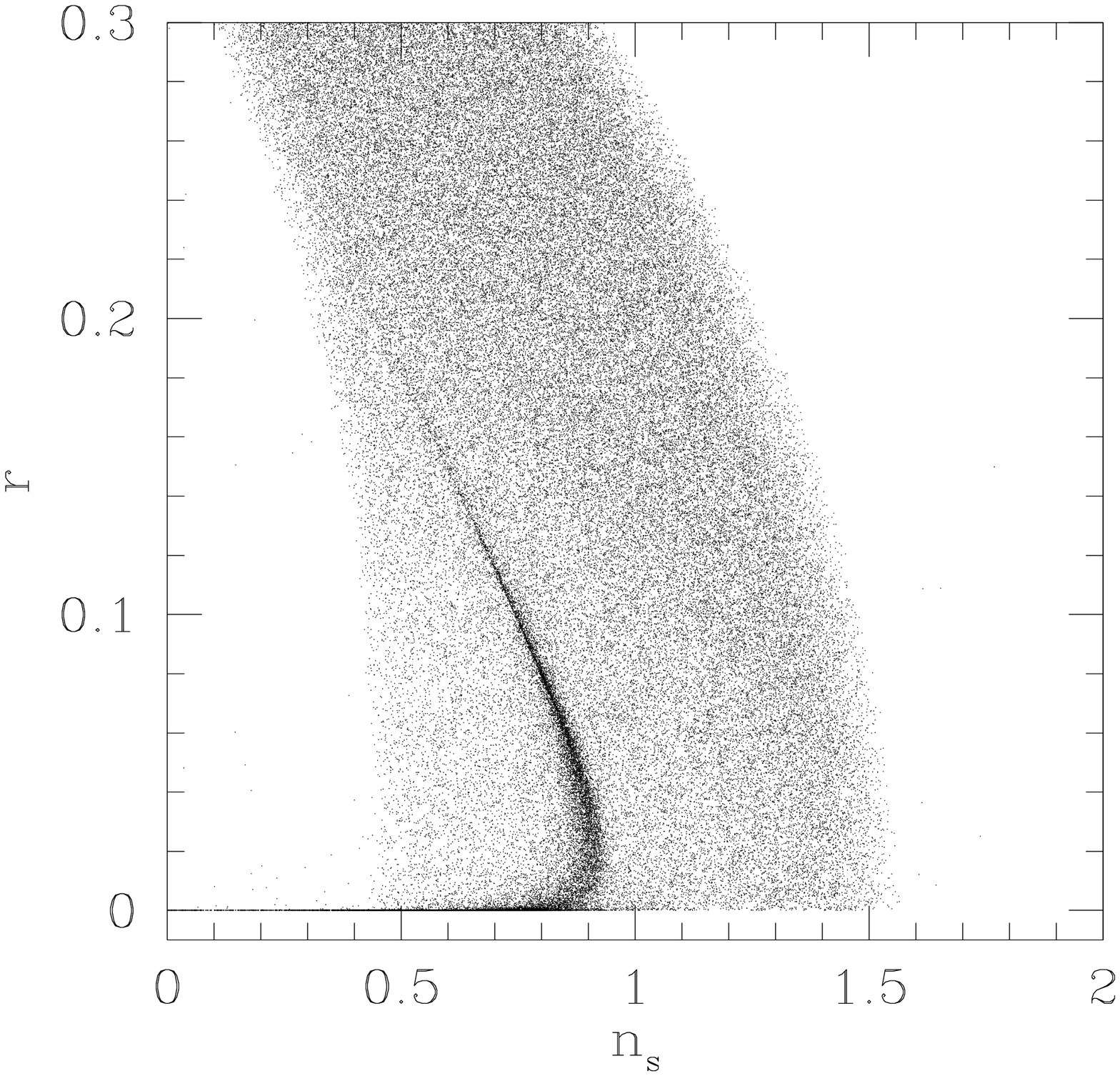}
\includegraphics[width=4.1cm]{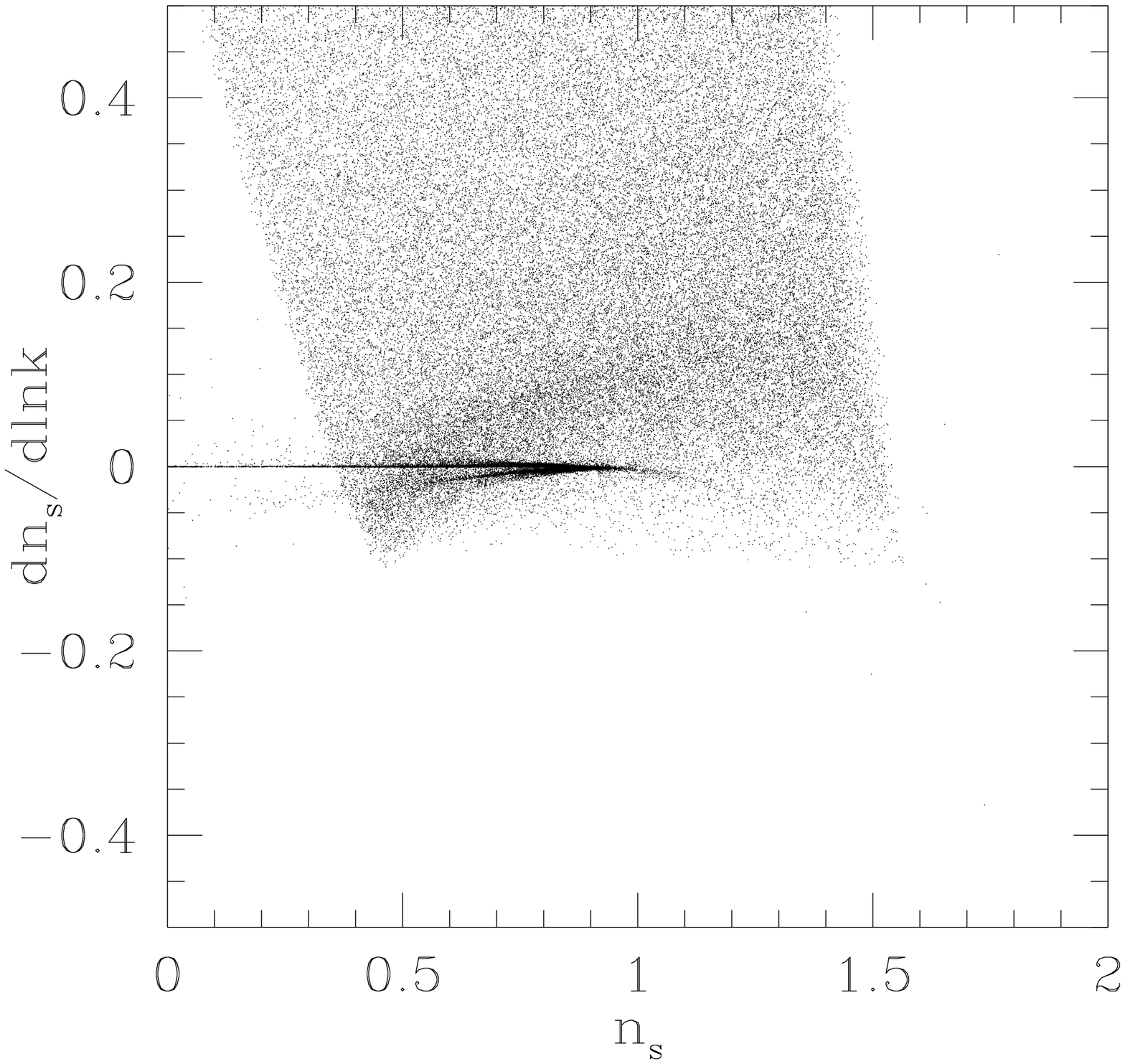}
\includegraphics[width=4.1cm]{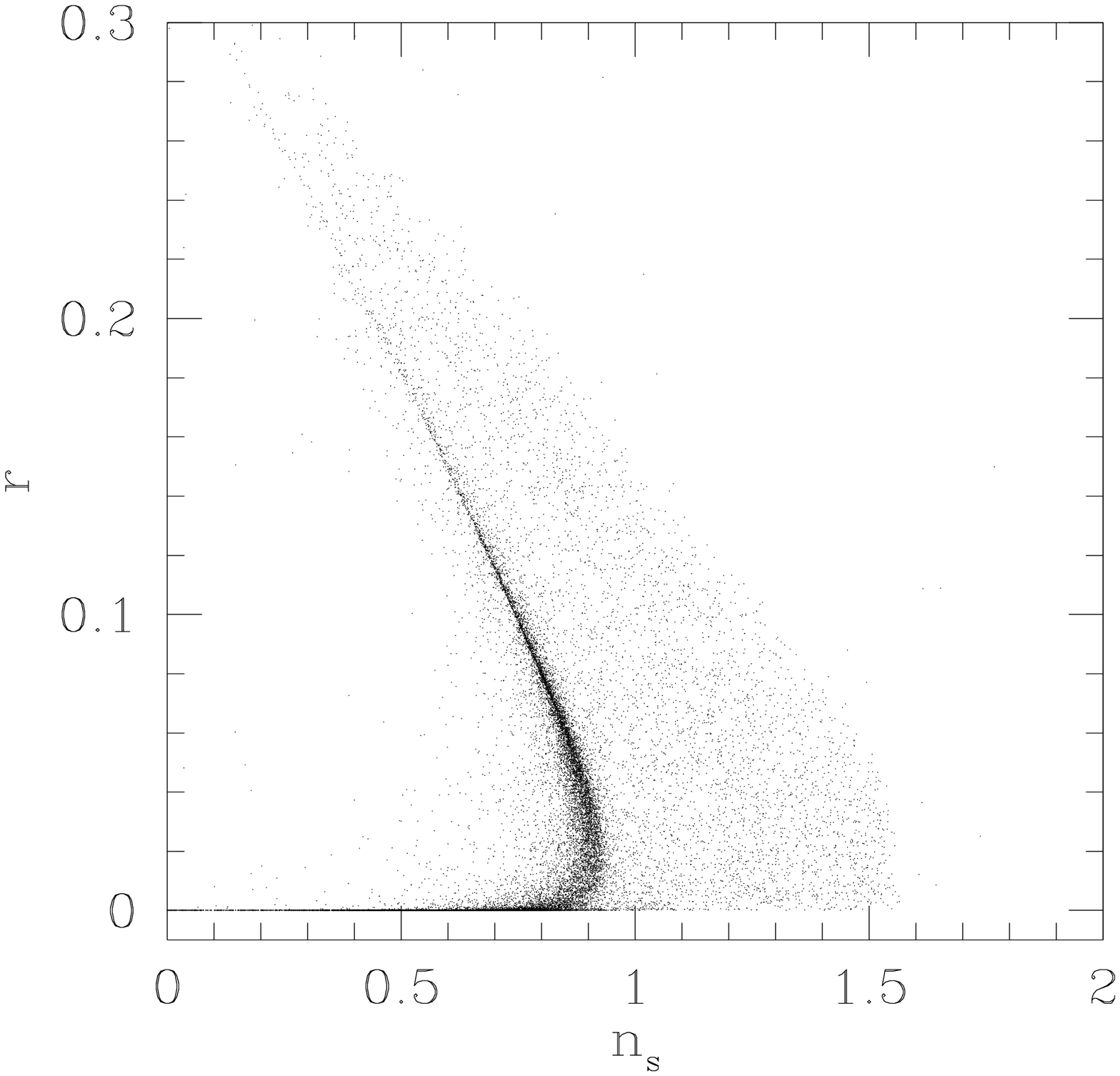}
\includegraphics[width=4.1cm]{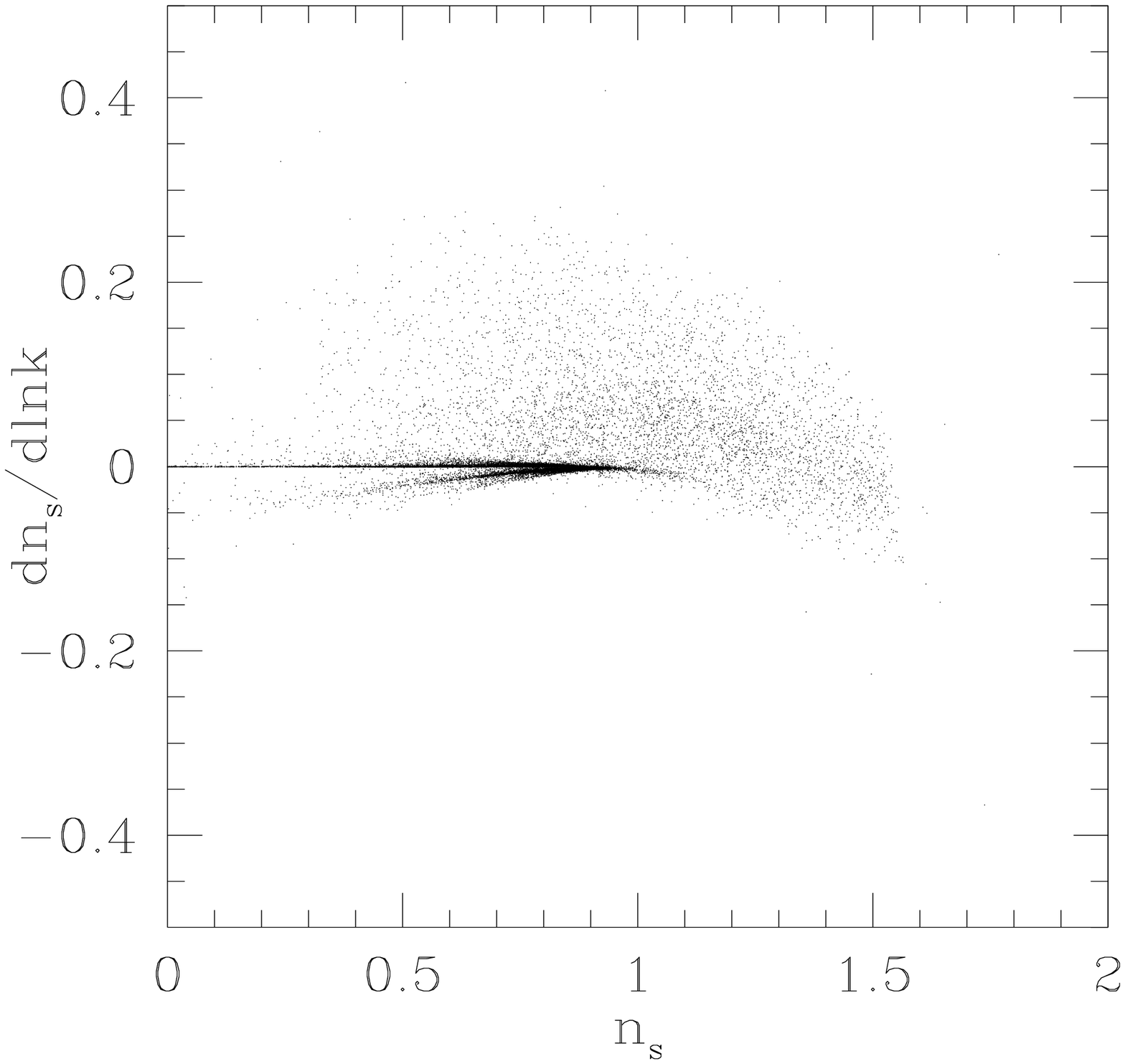}
\caption{The cosmological observables,
$n_{\rm s}$, ${\rm d} n_{\rm s} /{\rm d} {\rm{ln}} k$ and $r$, for models 
generated using the modified flow algorithm described in the text.  All models which sustain the required number of e-foldings of inflation are shown in the top two plots and those remaining once PBH constraints are applied are shown in the bottom plots.}
\label{stochasticresults}
\end{center}
\end{figure}

We use the modified flow algorithm described in sec.~\ref{flow} to
generate a large ensemble (250,000) of inflation models.  In
fig.~\ref{stochasticresults} (top row) we plot the cosmological observables 
for all models which are able to sustain the
required number of e-foldings of inflation, $N_{\rm{cos}}$.  In around $7\%$ of 
the models inflation ends naturally via $\epsilon_{\rm H}=1$ and these largely populate
the concentrated diagonal feature seen in the left hand plots as well as the $r=0$ line (c.f. Ref.~\cite{Kinney:2002qn}). In the remaining $93\%$ of models, $\epsilon_{\rm{H}} \rightarrow 0$ and it is
assumed, c.f. Ref.~\cite{Peiris:2008be}, that a secondary mechanism,
such as hybrid inflation, acts to end inflation in these cases. Large
positive running is in principle allowed (see top right plot), however these models may
have large amplitude perturbations on small scales and hence
overproduce PBHs.

To apply the PBH constraints we use the Stewart-Lyth expression for
the power spectrum, eq.~(\ref{stewartlyth}), to identify inflation
models where the amplitude of the perturbations on small scales which
exit the horizon close to the end of inflation is large, and may lead
to the over-production of PBHs.  For these models, we then carry out
an accurate numerical evolution of the primordial perturbations, as
described in Sec.~\ref{pert}.

The PBH abundance constraints have recently been compiled and updated
in Refs.~\cite{Josan:2009qn,Carr:2009jm}. The resulting constraints on the
amplitude of the power spectrum are typically in the range
$\mathcal{P_{\mathcal{R}}} < 10^{-2}-10^{-1}$ with some scale
dependence~\cite{Josan:2009qn}.  To be conservative we use the
constraint $\mathcal{P_{\mathcal{R}}} < 10^{-1}$.
The bottom row of fig.~\ref{stochasticresults} shows the cosmological
observables for the models which remain once those which over-produce
PBHs are excluded.  The $7\%$ of the original models for which
inflation ends naturally generally have $n_{\rm s} < 1$ on all scales
and so are unaffected by the PBH constraints.  Of the remaining
models, in which inflation continues indefinitely in the absence of a
secondary mechanism, $92\%$ are excluded by PBH overproduction.  Of
the models initially generated, only approximately $1\%$ end via a
secondary mechanism and do not overproduce PBHs.  With an accurate
numerical calculation of the perturbations the number of these models
decreases by approximately $10\%$.
Large positive running is now excluded as expected (see bottom-right plot).

Cosmological constraints on ${\rm d} n_{\rm s} /{\rm d} {\rm{ln}} k$
eliminate a significant fraction of the models
generated using flow algorithms~\cite{Kinney:2002qn}.  A full MCMC
analysis of cosmological data is beyond the scope of this work,
however a simple application of the observational constraints shows
that a significant fraction of cosmologically viable models are
excluded by PBH constraints.  Of the models generated using our
modified flow analysis which have cosmological observables within the
3$\sigma$ ranges found by WMAP7~\cite{Komatsu:2010fb} $19\%$ are 
excluded by PBH over-production. This illustrates
that in the era of precision cosmological measurements PBH still
provide a powerful constraint on inflation models.

\section{Conclusions}
\label{discuss}

We have applied constraints on the primordial power spectrum from the
overproduction of primordial black holes to inflation models generated
by a modified flow algorithm. The amplitude of inflationary
perturbations is usually calculated using the
Stewart-Lyth~\cite{Stewart:1993bc} expression, however for scales
which exit the horizon close to the end of inflation the assumptions
underlying this expression are violated. A numerical calculation is
therefore required, and the amplitude of the perturbations on small
scales can be significantly enhanced~\cite{Leach:2000yw,Bugaev:2008bi}.
The models generated by the modified flow algorithm which end naturally (roughly
$7\%$ of the total) generally have a red spectrum of perturbations on
all scales and so are unaffected by PBH constraints.
The remaining $93\%$ of models
equations have a late time attractor with $\epsilon_{H} \rightarrow
0$ and it is assumed that an auxiliary mechanism
terminates inflation. The majority of these models are however excluded due to PBH over-production. The number of viable models decreases
if the power spectrum is calculated numerically.
Of the models generated using our modified flow analysis which have 
cosmological observables within the 3$\sigma$ ranges found by 
WMAP7~\cite{Komatsu:2010fb} $19\%$ are excluded by PBH over-production.

We conclude that PBH constraints provide a significant constraint on
models of inflation. Furthermore to exploit their full power an
accurate numerical calculation of the amplitude of primordial perturbations 
on small scales, which exit the horizon close to the end of inflation, is
required.
\begin{acknowledgments}
We are grateful to Andrew Liddle and Will Hartley for
useful discussions and acknowledge the use of the Inflation v2
module (written by Julien Lesgourgues and Wessel Valkenburg). AJ is
supported by the University of Nottingham, AMG is supported by STFC.
\end{acknowledgments}


%\bibliography{pbhinfsub3}
%\bibliographystyle{aa}

\providecommand{\href}[2]{#2}\begingroup\raggedright\endgroup

\end{document}